\documentclass[12pt]{article}
\usepackage[utf8]{inputenc}
\usepackage{fullpage}
\usepackage{graphicx}
\usepackage{amsmath}
\usepackage{amssymb}
\usepackage{float}
\usepackage{mathtools}
\usepackage[font=small,labelfont=bf]{caption}

\usepackage{listings}
\usepackage{color} 
\definecolor{mygreen}{RGB}{28,172,0} 
\definecolor{mylilas}{RGB}{170,55,241}
\usepackage{pdfpages}
\DeclarePairedDelimiter\abs{\lvert}{\rvert}%

\title{Mode Division Multiplexing (MDM) Weight Bank Design for Use in Photonic Neural Networks}
\author{Ethan Gordon '17 \\ \em{egordon@princeton.edu} \\ \\ Advisor: Paul R. Prucnal \\ \em{prucnal@princeton.edu} \\ \\ Submitted in partial fulfillment \\ of the requirements for the degree of \\ Bachelor of Science in Engineering \\ Department of Electrical Engineering \\ Princeton University \\ \\}
\date{May 10, 2017}

\begin{document}

\lstset{language=Matlab,%
    breaklines=true,%
    morekeywords={matlab2tikz},
    keywordstyle=\color{blue},%
    morekeywords=[2]{1}, keywordstyle=[2]{\color{black}},
    identifierstyle=\color{black},%
    stringstyle=\color{mylilas},
    commentstyle=\color{mygreen},%
    showstringspaces=false,
    numbers=left,%
    numberstyle={\tiny \color{black}},
    numbersep=9pt, 
    emph=[1]{for,end,break},emphstyle=[1]\color{red}, 
}

\maketitle

\newpage

\section*{Honor Statement}
I hereby declare that this Independent Work report represents my own work in accordance with University regulations.
\begin{flushright}
\includegraphics[width=0.2\textwidth]{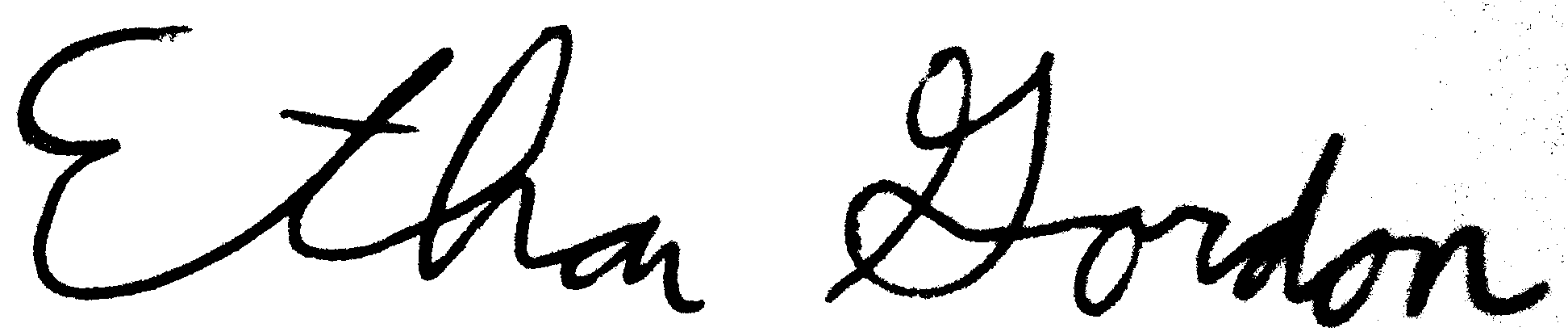} \\
Ethan K. Gordon '17
\end{flushright}

\newpage
\begin{centering}
{\LARGE Mode Division Multiplexing (MDM) Weight Bank Design for Use in Photonic Neural Networks} \\
\vspace*{10px}
{\large Ethan Gordon '17} \\
{\em egordon@princeton.edu} \\
\end{centering}

\section*{Abstract}
Neural networks provide a powerful tool for applications from classification and regression to general purpose alternative computing. Photonics have the potential to provide enormous speed benefits over electronic and software networks, allowing such networks to be used in real-time applications at radio frequencies. Mode division multiplexing (MDM) is one method to increase the total information capacity of a single on-chip waveguide and, by extension, the information density of the photonic neural network (PNN). This Independent Work consists of three experimental designs ready for fabrication, each of which investigates the process of expanding current PNN technology to include MDM. Experiment 1 determines the optimal waveguide geometry to couple optical power into different spacial modes within a single waveguide. Experiment 2 combines MDM and previous wavelength division multiplexing (WDM) technology into a single weight bank for use as the dendrite of a photonic neuron. Finally, Experiment 3 puts two full neurons in a folded bus, or "hairpin," network topology to provide a platform for training calibration schemes that can be applied to larger networks.
\newpage

\section*{Acknowledgements}
First and foremost, I would like to acknowledge Alex Tait, the graduate student that guided me through all the stages of this project. I am excited to continue working with Alex next year for my Senior Independent Work. I would also like to extend my thanks to everybody in the Lightwave Communications Laboratory, from graduate students and post-doctoral researchers to Prof. Prucnal, for their continued support and advice. Finally, the fabrication of all experiments would not have been possible without funding from the School of Engineering and Applied Science and the Department of Electrical Engineering. Thank you for making my Independent Work a success. \\

\noindent {\em This work is dedicated to ELE Car Lab 2016 and all the late nights we spent together.} 
\tableofcontents

\newpage
\section{Motivation and Background}
Neural networks are a continued staple of machine learning and alternative computing, with applications ranging from classification and regression to general-purpose computation. The general concept is motivated by biological nervous systems. Networks are composed of individual neurons, each of which takes as input a weighted sum of channels from the rest of the network, applies some non-linear function (commonly a threshold or logarithm), and outputs the result on its own channel back into the network. Networks in general can be modeled in software, but can and have been reproduced physically to tackle fast real-time problems. It is in these fast real-time systems that Photonic Neural Networks (PNNs) can have the greatest impact.
\subsection{The Size-Bandwidth Trade-off}

\begin{figure}[h!]
\centering
\includegraphics[width=0.5\textwidth]{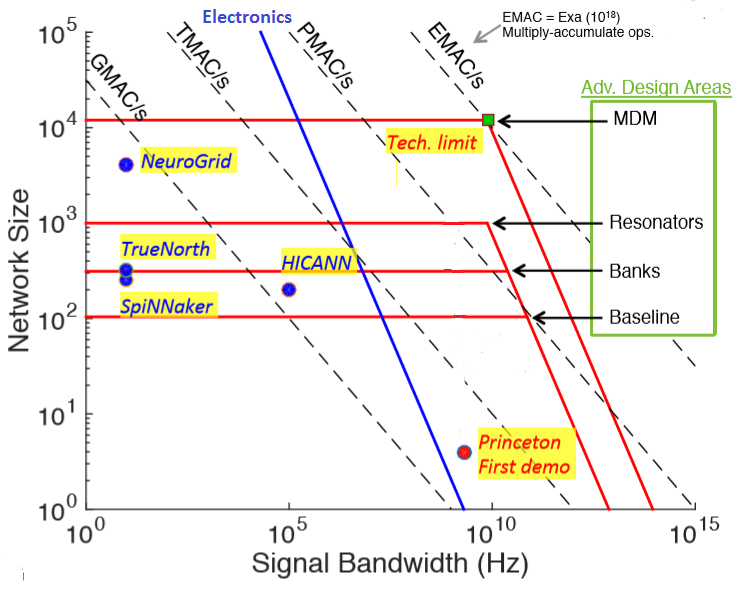}
\caption{Trade-off between Network Size (in number of channels, determined in photonics by the channel selectivity of the weight banks) and each channel's Bandwidth, the product of which is the total information capacity of the network. Current electronic networks are shown in blue text, with the electronic bandwidth-size trade-off also in blue. Red lines show trade-offs for existing or proposed technologies in photonic networks. Photonic channel limits from \cite{mrr}.}
\label{fig:bandwidth}
\end{figure}
One fundamental limit on neural networks is the tradeoff between network size and the total signal bandwidth. As more neurons go on the network, it takes longer for information to transverse the network, either due to the speed of the network or some limit on how much information the network can transmit at any given time. Figure \ref{fig:bandwidth} demonstrates the current state of this trade-off in electronic and photonic systems. Even one of the faster electonic systems is bandwidth limited to 100kHz, with that limit decreasing as the network grows into multiple wafers due to inter-wafer communication delay \cite{hicann}. In general, while electronic neural networks can contain tens of thousands of neurons, the slower speeds limit these networks to trillions of operations per second (TMAC/s) or slower. Photonic networks have the potential to break this barrier, without appreciably increasing network size.

\begin{figure}[h!]
\centering
\includegraphics[width=0.5\textwidth]{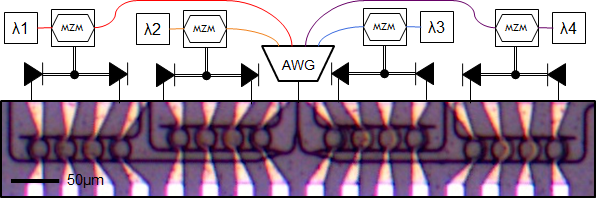}
\caption{Princeton's first photonic network demonstration.}
\label{fig:ppnn}
\end{figure}

Princeton recently demonstrated its first working physical photonic neural network, as shown in Figure \ref{fig:ppnn}, containing four neurons on different wavelengths \cite{demo}. The network utilizes a {\em star topology}, where the outputs of all the neurons enter the network at a single point, splitting off from that point to meet the weight banks. Each weight bank is a collection of Microring Resonators (MRR). Each MRR is tuned to a different wavelength, with an electric heater deforming the ring ever so slightly to adjust the actual percentage of light that couples through the ring, effectively applying a weight to that wavelength channel independent of other channels. This method for weighting each channel of a wavelength division multiplexed (WDM) signal has been the principal passive optical component in all Princeton Neural Network designs \cite{mrr}. Once weighted, the addition occurs in a pair of balanced photodiodes, allowing for both positive and negative weights. The electric signal immediately goes to a Mach-Zehnder Modulator, whose non-linear properties act as the axon of the neuron itself. Each neuron modulates a different wavelength, and all wavelengths are multiplexed off-chip and fed back into the point of the star network.

While Princeton's demonstration is the first of its kind, it does suffer from some drawbacks. First of all, it is only four neurons large. The star topology has trouble scaling on-chip, due to the need for every point on the network to have a direct connection to the center of the star. Also, the large amount of off-chip fiber did slow the device into the kilohertz range, though that problem will not exist in a fully integrated system. Finally, in the long term, there are limits to both the channel selectivity of the microring resonators (channels per bandwidth) and the overall bandwidth of the waveguide, the product of which determines the total number of channels available in a network. \\

This work investigates the use of mode division multiplexing (MDM) to expand the information capacity of photonic neural networks by an order of magnitude.

\subsection{Current Work in MDM}

\begin{figure}[h!]
\centering
\includegraphics[width=0.5\textwidth]{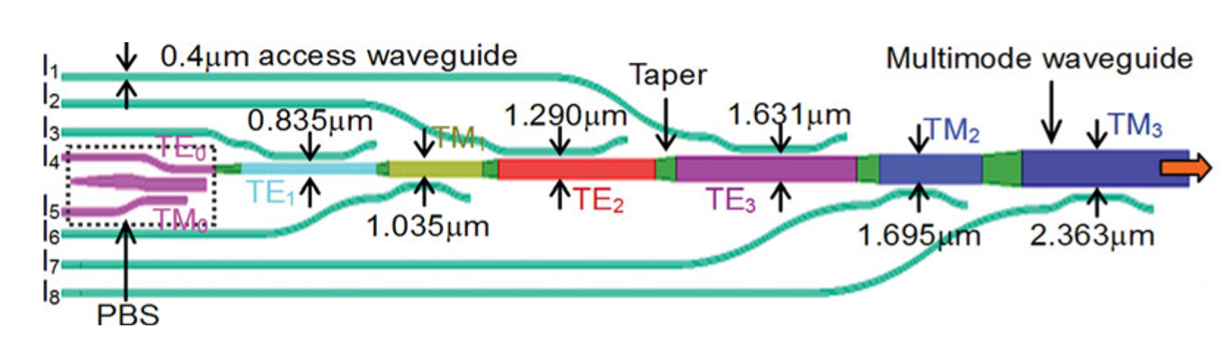}
\caption{Existing design for an on-chip mode (de)multiplexer \cite{onchip}. Coupling occurs at asymmetric Y-junctions with varying bus widths.}
\label{fig:muxdemux}
\end{figure}

Previous work on MDM in silicon has focus solely on the actual multiplexer / demultiplexer. The general strategy is to couple signals on a single-mode waveguide to the highest order mode on a multi-mode waveguide, as shown in Figure \ref{fig:muxdemux} and demonstrated in silicon \cite{onchip}. After each coupling portion for the Nth order mode, the bus waveguide adiabatically tapers larger to couple the (N+1)th order mode. The adiabatic taper, which occurs gradually, prevents any intermodal mixing between coupling. An exact mirror image of the device provides demultiplexing.

\begin{figure}[h!]
\centering
\includegraphics[width=0.5\textwidth]{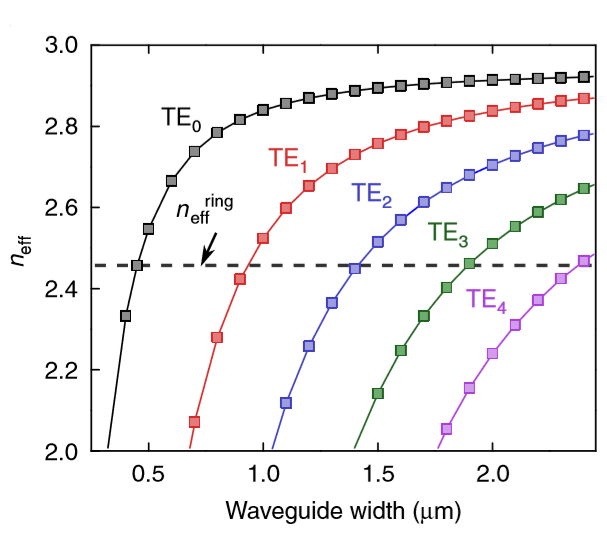}
\caption{Example of simulated effective index data for waveguides of differing widths \cite{wdm}. Optimal coupling occurs when the effective index of the multi-mode waveguide matches that of the single-mode MRR (example shown as dotted line).}
\label{fig:neff}
\end{figure}

The principle behind the selectivity of the mode coupling is that the effective index in a waveguide for a given mode is dependent on the width, as shown in example simulation data in Figure \ref{fig:neff} and demonstrated in silicon \cite{wdm}. Optimal coupling occurs between two modes in two different waveguides when the effective indices match, and so if a given width is optimal for coupling to a given mode, it will be far from optimal for other modes. In practice, it is easiest to set the width of a multi-mode waveguide such that its highest order mode will couple to the single-mode MRR, as that mode already determines the minimum width of the waveguide, and then once that mode is coupled out, taper to the width for coupling with the next-highest-order mode.

\section{Experiment 1: Waveguide Geometry}
The goal of Experiment 1 is to determine the optimal coupling geometry between single- and multi-mode waveguides, since the value of the effective index differs by material and wavelength, and could diverge significantly from simulated data.
\subsection{Experimental Design}
\begin{figure}[h!]
\centering
\includegraphics[width=0.5\textwidth]{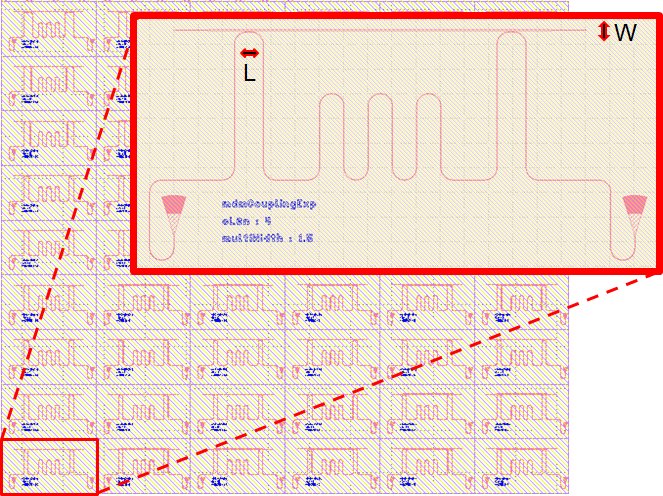}
\caption{Asymmetric Mach-Zehnder Interferometer. The goal is to determine optimal waveguide width (W) and coupling length (L) for 100\% coupling.}
\label{fig:mz}
\end{figure}

\noindent The design of this experiment is an asymmetric Mach-Zehnder interferometer. Given an input waveform $E_{in}$, the transfer function of the device can be written as:
\begin{equation}
\begin{bmatrix}
    E_{out} \\
    E_{taper} \\
\end{bmatrix} = 
M \begin{bmatrix}
    E_{in} \\
    0 \\
\end{bmatrix}
\end{equation}
Where M is a composition of both single-mode to multi-mode couplers and a phase shift $\Delta\phi$ due to the difference in optical length of the two paths through the interferometer.
\begin{equation}
M = M_{coupler} * M_{\Delta\phi}*M_{coupler}
\end{equation}
$M_{coupler}$ can be determined given a power coupling ratio $\alpha$, conservation of energy, and a $\frac{\pi}{2}$ phase shift that occurs as power couples from one waveguide to another. 
\begin{equation}
M_{coupler}=
\begin{bmatrix}
    \sqrt{1-\alpha} & j\sqrt{\alpha} \\
    j\sqrt{\alpha} & \sqrt{1-\alpha} \\
\end{bmatrix}
\end{equation}

\noindent Combined with the phase shift $\Delta\phi=k*\Delta L$, where k is the wavenumber related to the wavelength of the light, we can determine the transfer coefficient between the input waveform and the measured output:
\begin{equation}
M = 
\begin{bmatrix}
    \sqrt{1-\alpha} & j\sqrt{\alpha} \\
    j\sqrt{\alpha} & \sqrt{1-\alpha} \\
\end{bmatrix}
\begin{bmatrix}
    e^{jk\Delta L} & 0 \\
    0 & 1 \\
\end{bmatrix}
\begin{bmatrix}
    \sqrt{1-\alpha} & j\sqrt{\alpha} \\
    j\sqrt{\alpha} & \sqrt{1-\alpha} \\
\end{bmatrix}
\end{equation}
\begin{equation}
\abs*{\frac{E_{out}}{E_{in}}}^{2} = \abs*{(1-\alpha)e^{jk\Delta L} - \alpha}^{2} = \alpha^{2} + (1-\alpha)^{2} - 2\alpha(1-\alpha)cos(k\Delta L)
\end{equation}
\noindent Therefore, a power spectrogram of the device will result in a cosine with a DC-shift. The amplitude of that cosine, called the {\em Extinction Ratio}, is only dependent on the coupling ratio between the single-mode and multi-mode waveguides.
\subsection{Simulations and Expected Results}
\begin{figure}[h!]
\centering
\includegraphics[width=0.8\textwidth]{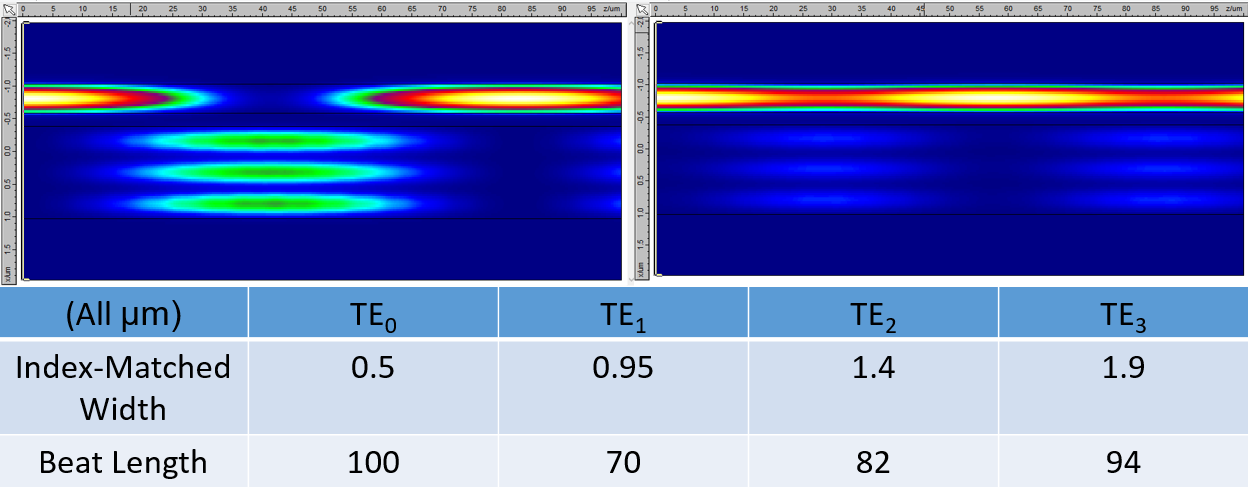}
\caption{Simulation results for coupling between the fundamental mode ($TE_{0}$) and the second order mode ($TE_{2}$), for both the index-matched width (left) an a width mismatched by 20nm (right). Note that optical power oscillates back and forth between the waveguides over the course of one {\em beat length}. Below is a table of simulated index-matched widths and beat lengths for each mode.}
\label{fig:simres}
\end{figure}

We can use simulations of optical coupling between single-mode and multi-mode waveguides to determine the relationship between W, L, and the coupling ratio $\alpha$. Some of these results are presented in Figure \ref{fig:simres}. As shown previously, optimal coupling happens when the width of the multi-mode waveguide is such that the indices of refraction in the two waveguides match. Even a slight mismatch can drastically decrease the coupling ratio. Once the indices are matched, optical power oscillates back and forth between the two waveguides. The spacial period of this oscillation is called the {\em beat length}, and differs by mode. Optimal coupling occurs at half of the beat length, when all power has moved into the multi-mode waveguide. 
\begin{equation}
\alpha = f(W)*(1 - cos(2\pi\frac{L}{(beat\ length)}))/2
\end{equation}

The first step in the experiment is to maximize the extinction ratio, which corresponds with $\alpha=\frac{1}{2}$. This implies that the width matches effective index and the coupling length is one-quarter or three-quarters of the beat length. The next step is to minimize the extinction ratio {\em without} changing the width. Since the extinction ratio is minimized with both $\alpha=1$ {\em and} $\alpha=0$, the simulation data can be used to help distinguish the two cases. It should be clear whether the coupling length is closer to the simulated beat length or the simulated half-beat length.

\section{Experiment 2: Weight Bank}
\subsection{Experimental Design}
\begin{figure}[h!]
\centering
\includegraphics[width=0.8\textwidth]{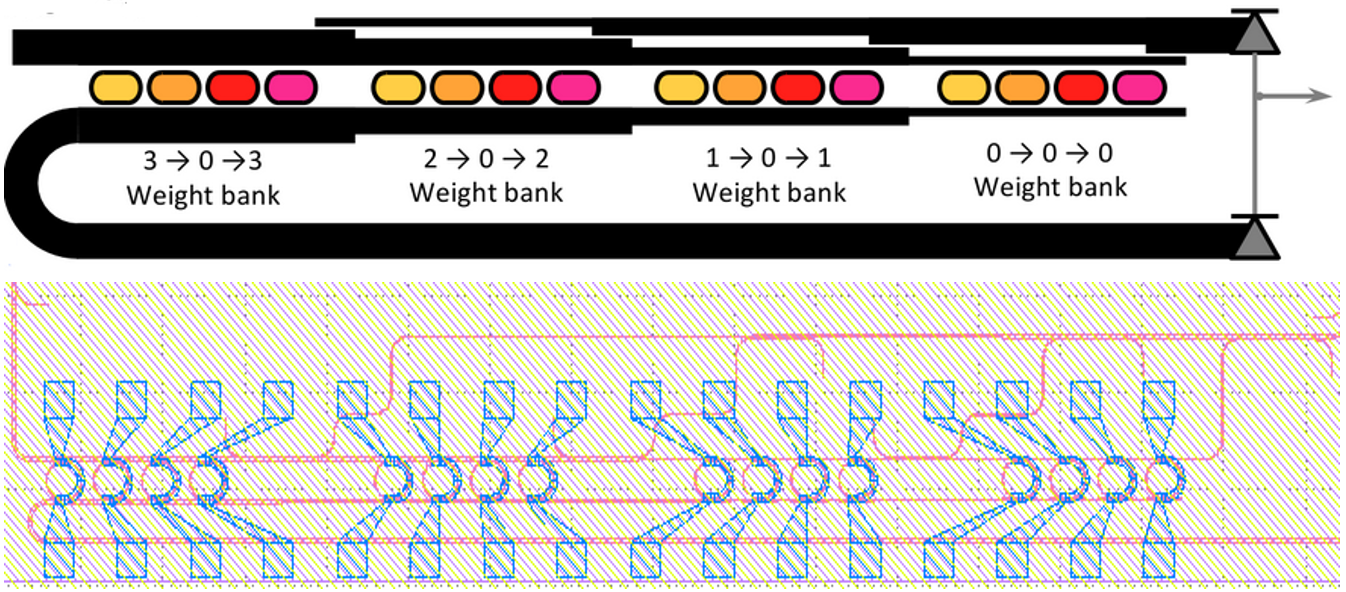}
\caption{(Top) Schematic for a four-mode four-wavelength MDM/WDM weight bank. (Bottom) Layout of the weight bank on-chip. Electric heaters are shown in blue.}
\label{fig:wb}
\end{figure}

Using the simulated optimal coupling geometry, we designed a combined MDM / WDM weight bank to perform the weighted addition on the dendrite side of the neuron, as shown in Figure \ref{fig:wb}. The bank is composed of a series of stages, one per mode. Each stage is similar to th WDM weight bank from Princeton's network demonstration: a series of MRRs tuned to each wavelength channel in the waveguide. Each MRR transmits a certain percentage of the optical power in its specific wavelength channel from the input bus into the "drop" waveguide. The exact percentage is controlled by electric heaters that can slightly modify the exact perimeter of the MRR. The rest of the light in the channel is coupled completely into the "add" waveguide. Finally, the input waveguide is adiabatically tapered down to the coupling width of the next-highest-order mode for the next stage, until the final stage has only a single-mode waveguide.

Both the "add" and "drop" waveguides are fed into two balanced photodiodes, which provides summation by converting the optical power into an electric current. The difference in current between the two photodiodes is the current that acts as the input for the neuron itself. The benefit of using the photodiode for summation is that it is mode- and wavelength-agnostic. Based on the mechanics of multi-mode coupling, each mode in the input gets coupled into a different mode of the "add" waveguide, However, since the {\em total} optical power in that waveguide is still a correct weighted sum of each channel, the correct electrical current is drawn by the photodiode.

\subsection{Intermodal Mixing}
The summation in the photodiode also solves another issue with the weight bank: intermodal mixing. When there is a bend in a multi-mode waveguide, the mode basis changes throughout the course of the bend. The result is a shuffling of optical power between modes. Assuming no bending loss, this shuffling can be accurately modeled as a linear transformation with a unitary matrix.
\begin{equation}
\begin{bmatrix}
    TE_{0}' \\
    TE_{1}' \\
\end{bmatrix}
= M \begin{bmatrix}
    TE_{0} \\
    TE_{1} \\
\end{bmatrix}
\end{equation}
\begin{equation}
M^{\dag}M = 1
\end{equation}

\noindent Since the transformation is unitary, the total optical power in the waveguide does not change. Therefore, within the wavebank itself, the bend in the "drop" waveguide makes no difference in the summation done by photodiode.

\begin{figure}[h!]
\centering
\includegraphics[width=0.5\textwidth]{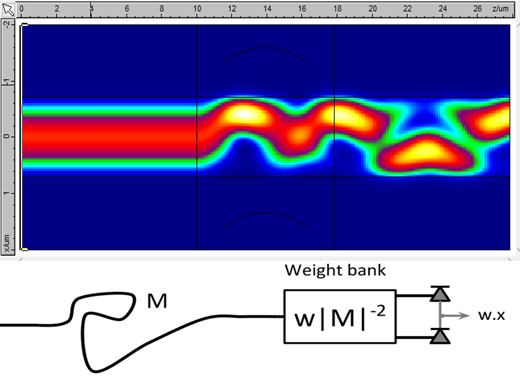}
\caption{(Top) Simulation of the fundamental mode of a two-mode waveguide going through a bend. The result is a mix of the fundamental and first-order mode. (Bottom) How a weight bank can undo the effects of an intermodal mixing matrix.}
\label{fig:intermode}
\end{figure}

Closer inspection of Figure \ref{fig:wb} would reveal that intermodal mixing also occurs outside of the weight bank in the form of a bending input waveguide. This could not be done in previous muilti-mode experiments, as there is no easy way to undo the mode mixing on chip through waveguide geometry alone. Both \cite{onchip} and \cite{wdm} use only straight multi-mode waveguides. Fortunately, a properly calibrated weight bank can be used to undo all intermodal mixing in the network. A weighted summation is mathematically equivalent to an inner product with a weighting vector:
\begin{equation}
    y = w \cdot \abs{x}^{2}
\end{equation}
\noindent where $\abs{x}^{2}$ is the vector of optical power in each channel. If x is modified by a mix matrix $\abs{M}^{2}$, then the same output can be preserved if w is modified by M's inverse.
\begin{equation}
    w\abs{M}^{-2} \cdot \abs{M}^{2}\abs{x}^{2} = w \cdot \abs{x}^{2} = y
\end{equation}
\noindent By experimenting with known input vectors x and weights w, the individual components of M can be measured, allowing for the calibration of the weight bank. Calibrating each weight bank on the network allows for arbitrary bending in any multi-mode waveguide in the network, leading to more flexible network design and the ability to fit more neurons on a smaller surface area.
\section{Experiment 3: Multi-Neuron Network}
\subsection{Experimental Design}
\begin{figure}[h!]
\centering
\includegraphics[width=0.3\textwidth]{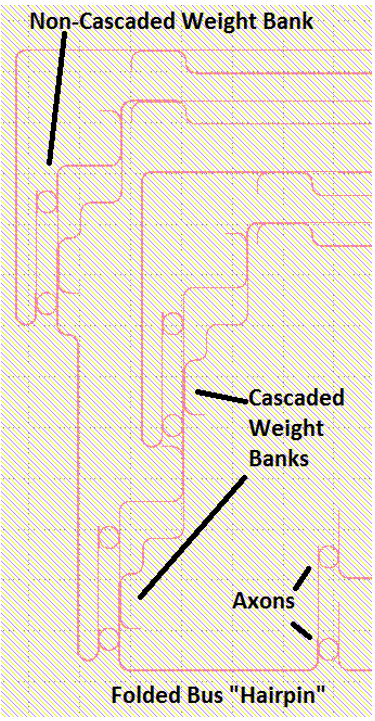}
\caption{Final CAD of a two-neuron network, including axons, one cascaded weight bank, and one non-cascaded weight bank.}
\label{fig:network}
\end{figure}

The next step is to combine muiltiple weight banks into a full network, as shown in Figure \ref{fig:network}. This design has two neurons on two channels on two different modes and one wavelength. Optical power is pumped into the network to the axons, which are just MRRs tied to electric heaters. Modulation of optical power via the electric heaters follows a Lorentzian function, providing the non-linear element that can act as a continuous (non-spiking) neuron. Each neuron uses the same wavelength of light, but couples onto a different mode in the main bus. The bus wraps around into a series of weight banks. The first bank cuts the power in half, while the subsequent weight banks provide the actual weighting before the light goes off-chip to be summed and fed back into the axons.

\subsection{"Hairpin" Topology and Cascaded Banks}
\begin{figure}[h!]
\centering
\includegraphics[width=0.5\textwidth]{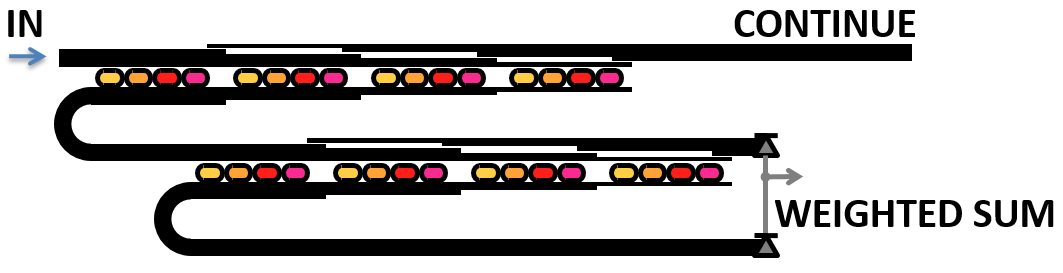}
\caption{Schematic of a cascaded weight bank. The first bank drops a fixed percentage of power from the bus, while the second performs the actual weighted addition.}
\label{fig:cascade}
\end{figure}

\begin{figure}[h!]
\centering
\includegraphics[width=0.5\textwidth]{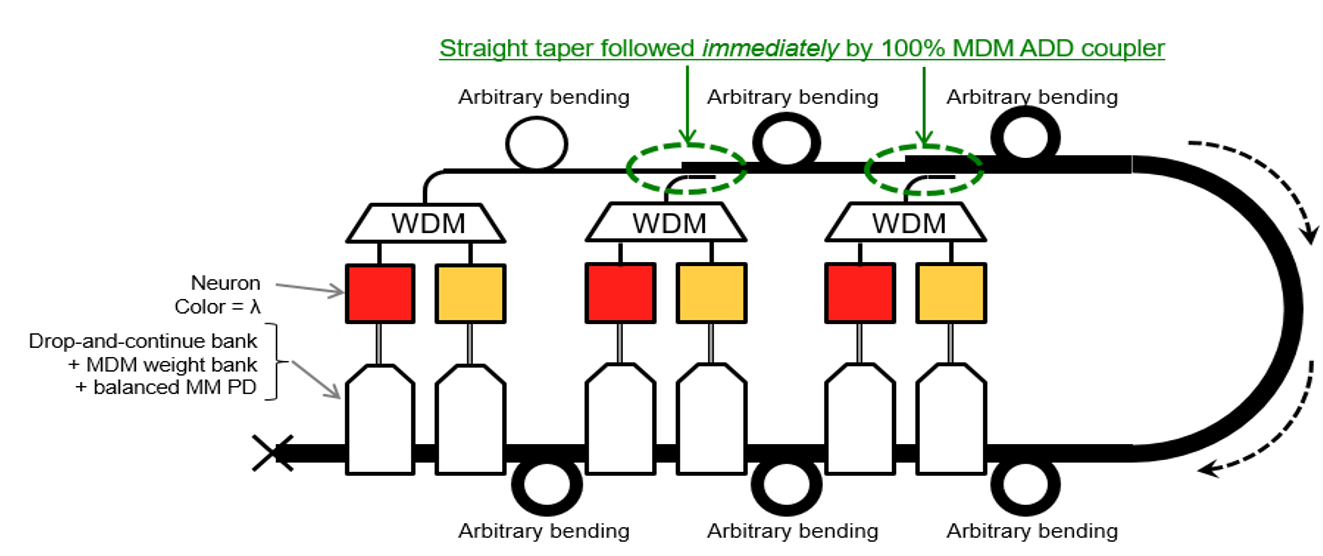}
\caption{Schematic of a folded bus, or "hairpin," network design. Arbitrary bending between neurons allows for a flexible on-chip network design.}
\label{fig:hairpin}
\end{figure}

As designed in Experiment 2, a weight bank has a single multi-mode input and two multi-mode outputs. To allow for multiple weight banks to share a single bus, and extra weight bank can be used with the sole purpose of dropping a fixed percentage of power from the shared bus before a second weight bank does the actual weighting. An example of these {\em cascaded weight banks} is shown in Figure \ref{fig:cascade}. While the extra weight bank does flip the mode channels between the input and the continue waveguides, this can just be factored into the mode mixing matrix at the next cascaded weight bank. In Experiment 3, the first neuron uses a cascaded weight bank, while the second (and last) neuron in the network just uses a single weight bank.
The actual layout of the network in Figure \ref{fig:network} is a folded bus, or "hairpin," topology. One limitation of the Princeton network demonstration is its limited ability to scale. Because every point in the star topology needs a direct connection to the source point, an addition of even one neuron can take up quite a bit of chip surface area. Furthermore, there is no easy way to split a multi-mode signal from one waveguide into multiple multi-mode waveguides. The hairpin topology places all axons on one side of the hairpin, and all dendrites on the other, as shown in Figure \ref{fig:hairpin}. Since there can be arbitrary waveguide bending between neurons, the hairpin can snake its way around the trip into an arbitrary configuration, allowing for flexible network and efficient use of chip surface area.
\section{Future Work and Applications}
The silicon chip with these three experiments will be fabricated during the summer, and the resulting data should be sufficient to design a larger network of neurons for high speed computations.
\subsection{Application: Digital Demixing}
\begin{figure}[h!]
\centering
\includegraphics[width=0.5\textwidth]{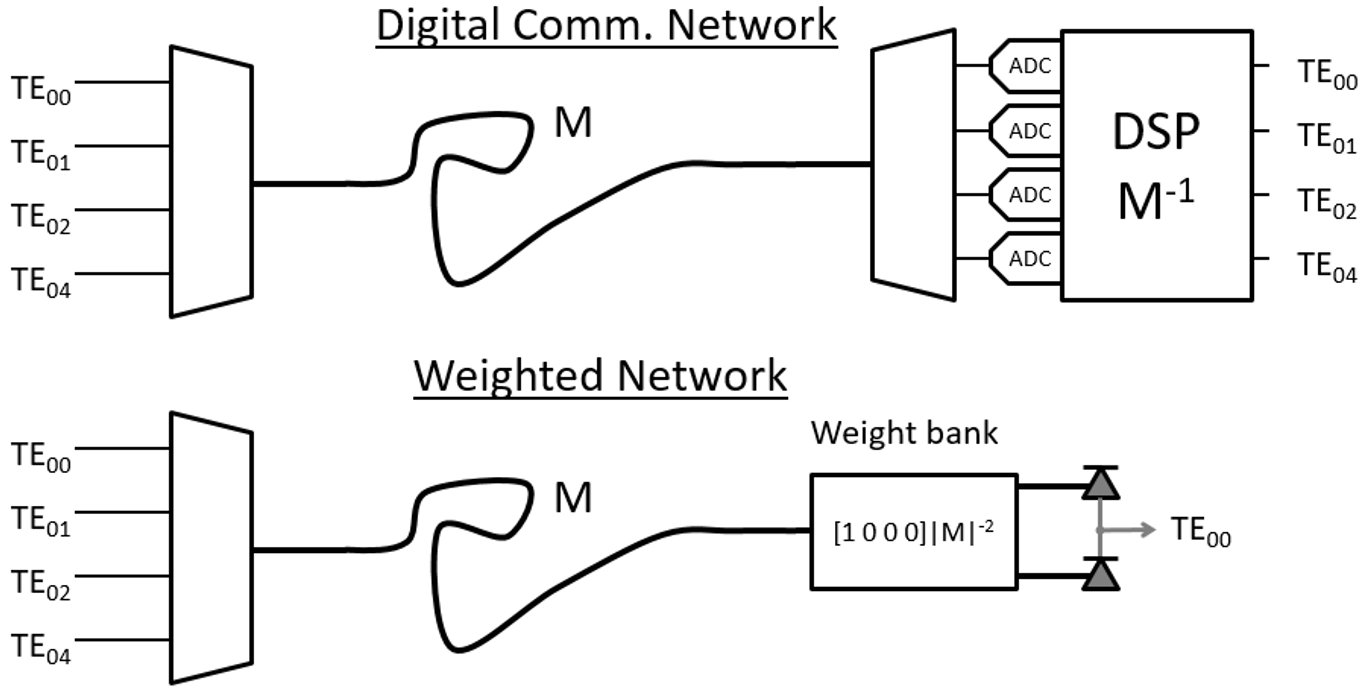}
\caption{Example of using a weight bank for channel selection. Multiple weight banks can completely demix all channels.}
\label{fig:demix}
\end{figure}
One application of the weight banks themselves is actually in digital applications. Multi-mode waveguides can greatly increase the aggregate bandwidth of digital optical systems on chip. However, intermodal mixing severely limits any bandwidth improvement, since DSP is required to invert the mixing matrix if there is any bending on chip. However, as shown in Figure \ref{fig:demix}, a single weight bank can be used to invert the mix matrix and select one of the mode channels. For N modes, N weight banks can be used in aggregate to completely invert the mixing matrix in the optical domain, without any DSP. 
\subsection{Application: RF Classification}
\begin{figure}[h!]
\centering
\includegraphics[width=0.5\textwidth]{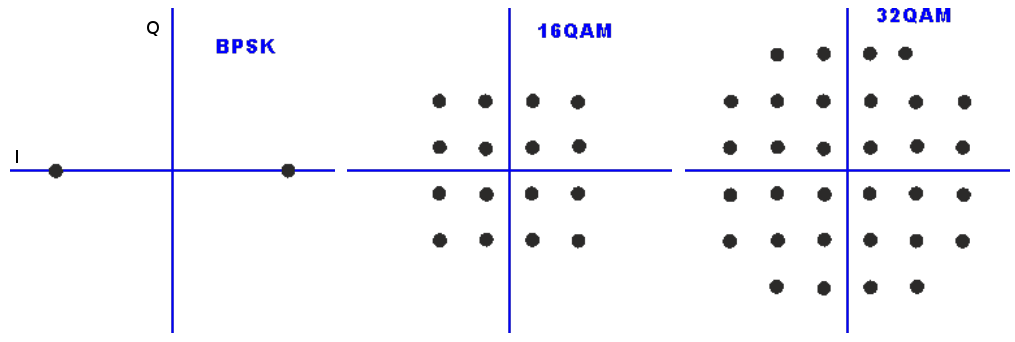}
\caption{Examples of I-Q digital modulation schemes. With more dots in the diagram, more data can be sent per "bit" interval.}
\label{fig:IQ}
\end{figure}
Finally, the full photonic neural network can be deployed for classification problems at Radio Frequencies. One practical example is modulation scheme classification. When broadcasting a signal over radio, one method of modulating digital data is to encode each bit as a combination of the amplitude of a cosine, the in-phase or I component, and the amplitude of a sin, the quadrature or Q component.
\begin{equation}
Y(t) = I(t)cos(\omega t) + Q(t)sin(\omega t)
\end{equation}
Examples of such schemes are shown in Figure \ref{fig:IQ}. A photonic neural network could take as inputs two analog parameters $I(t)$ and $Q(t)$, and output its best guess as to the modulation scheme in real time. Combined with flexible, software-defined radio, such networks could be used to sweep large swaths of the radio spectrum searching for data channels. \\

Photonic neural networks have the potential to be a driving force in high speed processing and computing. MDM can become a vital component in drastically increasing network size and aggregate bandwidth, allowing for more complex applications in the future.

\newpage
\section{Code Appendix}
\noindent The on-chip layout of each experiment was procedurally generated using MATLAB.
\subsection{Project Definitions}
\begin{lstlisting}
% MainProject
% Author : Ethan

clear; close all; clc; format long; format compact;
% clear classes; % this also clears breakpoints, so no bueno

% Get to the right directory
thisFile = mfilename('fullpath');
thisDir = thisFile(1:end-length(mfilename));
cd(thisDir);

% Make sure the right paths are added, and other project paths are removed
tp = cd; cd('../Functions - AlexUtil');
makePaths(tp);

% Project meta data
cad = struct('author', 'Ethan', ...       % project lead author or company
    'fab', 'UBC', ...                  % fabrication facility
    'process', 'Ebeam', ...          % fabrication process
    'run', '06', ...                  % run name
    'name', 'MDMCoupling', ...             % project name
    'layermap', 'UBC_map', ...            % layer map name
    'uunit', 1e-6, ...                     % CAD scale (1e-6 - > microns)
    'dbunit', 1e-9, ...                   % CAD database unit (1e-9 - > nm)
    'size', [10000, 10000], ...             % Floorplan dimensions
    'margin', struct('left', 100, 'right', 100, 'top', 100, 'bottom', 100), ...   % safety margin
    'v', 'v1');                           % version number
save('cadMeta.mat','cad');

%% ----- chip-wide variables -------
itPerRow = 4;
itStart = 0;
wgBias = 0.0;

% Default experiment information
defaultExp = struct('name','defaultName', ...
    'deviceModel', @device_blank, ...
    'setBuilder', @buildSet_arrayed, ...
    'buildInfo', struct('approAli','top','approSp','wg','approBend',10), ...
	'fixedArgs', {0}, ...
    'pInfo', struct('name','null','val',0,'axis','x'), ...
	'gc', struct('ref','TE1550_SubGC_neg31_oxide_hooked','flip',-1,'sp',127,'nPer',2), ... %no longer supports parameters, use references
    'wg', Waveguide(.5+wgBias,[1;0],10,10), ...
    'thermal',Wire(4,[11;0],10,20), ...
	'intercon',Wire(5,[12;0],10,20), ...
    'contactInfo',{0}, ...
    'rect', [200,500], ...
    'offs', [0,0]);
defaultExp.contactInfo = {[]};
defaultExp.fixedArgs = {[]};

contCounter = 0;
			
% ------------------------
%% Describe cell mdmCouplingExp
% ------------------------
someRads = [(.5:.5:5),(6:13)]; %[1,2,4,8,15];
pInfoModBend(1) = struct('name', 'cLen', ...
                       'val', [0,1,2,3,4,5,10,15,20], ...
                       'axis','y');
pInfoModBend(2) = struct('name', 'multiWidth', ...
                       'val', someRads(1:1), ...
                       'axis','x');

mdmCouplingExp = ReadOptions(defaultExp,'name','mdmCouplingExp', ... 
                               'pInfo',pInfoModBend, ...
                               'deviceModel',@device_MultiInterferometer, ...
							   'setBuilder',@buildExpSet_modular_duo, ...
							   'thermal',[], ...
							   'rect',[460,260]);
        
% ------------------------
%% Describe cell mdmWeightBankExp
% ------------------------

numModes = 4;
numWL    = 4;

mrr = Microring(.2,defaultExp.wg.layer,defaultExp.wg.dtype,'w',.5,'radius',10,'straightLength',2);
fb = FilterBank_v2(defaultExp.wg,defaultExp.thermal,mrr,'orderN',numWL,'dL',0.1,'busHold',8);

multiDeviceModel = @device_MDMWeightBank;

dParams = [numModes, numWL];

mdmWeightBankExp = ReadOptions(defaultExp,'name','mdmWeightBankExp', ...
	'deviceModel',@device_ModeMUXArr, ... 
	'contactInfo',{1}, ...
	'thermal',ReadOptions(defaultExp.thermal,'l1pad',18), ...
	'intercon',ReadOptions(defaultExp.intercon,'l1pad',18), ...
	'gc',ReadOptions(defaultExp.gc,'flip',-1, 'nPer', numModes * 3), ...
    'fixedArgs',{multiDeviceModel, dParams, fb}, ... % Multi-Mode Modular Device Model and Parameters
	'rect',[400,1000]);

% ------------------------
%% Describe cell mdmWeightBankSafeExp
% ------------------------

numModes = 4;
numWL    = 4;

fb2 = fb.copy;
fb2.orderN = numWL;
fb2.recalculate();

multiDeviceModel = @device_MDMWeightBankSafe;

dParams = [numModes, numWL];

mdmWeightBankSafeExp = ReadOptions(defaultExp,'name','mdmWeightBankSafeExp', ...
	'deviceModel',@device_ModeMUXArrSafe, ... 
	'contactInfo',{2}, ...
	'thermal',ReadOptions(defaultExp.thermal,'l1pad',18), ...
	'intercon',ReadOptions(defaultExp.intercon,'l1pad',18), ...
	'gc',ReadOptions(defaultExp.gc,'flip',-1, 'nPer', numModes * 3), ...
    'fixedArgs',{multiDeviceModel, dParams, fb2}, ... % Multi-Mode Modular Device Model and Parameters
	'rect',[400,1000]);

% ------------------------
%% Describe cell busNtwrkExp
% ------------------------

numModesNet = 2;
numWLNet    = 1;

multiDeviceModelNet = @device_MDMWeightBank;

dParamsNet = [numModesNet, numWLNet];

busNtwrkExp = ReadOptions(defaultExp,'name','busNtwrkExp', ...
	'deviceModel',@device_NtwrkArr, ... 
	'gc',ReadOptions(defaultExp.gc,'flip',-1, 'nPer', 5*numModesNet), ...
    'fixedArgs',{multiDeviceModelNet, dParamsNet}, ... % Multi-Mode Modular Device Model and Parameters
	'rect',[400,1000]);

% ------------------------
%% Describe cell busNtwrkInvExp
% ------------------------
numModesNet = 4;
numWLNet    = 4;

multiDeviceModelNet = @device_MDMWeightBank;

dParamsNet = [numModesNet, numWLNet];

fb3 = fb2.copy;
fb3.orderN = numWLNet;
fb3.recalculate();

busNtwrkInvExp = ReadOptions(busNtwrkExp,'name','busNtwrkInvExp', ...
	'contactInfo',{3}, ...
	'thermal',ReadOptions(defaultExp.thermal,'l1pad',18), ...
	'intercon',ReadOptions(defaultExp.intercon,'l1pad',18), ...
	'gc',ReadOptions(defaultExp.gc,'flip',-1, 'nPer', 5*numModesNet), ...
	'fixedArgs',{multiDeviceModelNet, dParamsNet, fb3, -1}, ... % Multi-Mode Modular Device Model and Parameters
	'deviceModel',@device_NtwrkArrInverted, ...
	'rect',[400,1000]);

% ------------------------
%% Describe cell busNtwrkWDMExp
% ------------------------
numModesNet = 1;
numWLNet    = 4;

multiDeviceModelNet = @device_MDMWeightBank;

dParamsNet = [numModesNet, numWLNet];

fb4 = fb2.copy;
fb4.orderN = numWLNet;
fb4.recalculate();

busNtwrkWDMExp = ReadOptions(busNtwrkInvExp,'name','busNtwrkWDMExp', ...
	'contactInfo',{4}, ...
	'gc',ReadOptions(defaultExp.gc,'flip',-1, 'nPer', 5*numModesNet), ...
	'fixedArgs',{multiDeviceModelNet, dParamsNet, fb4, -1}, ... % Multi-Mode Modular Device Model and Parameters
	'rect',[400,1000]);


%% Put the experiment options in an array (comment them to suppress)
expArray = mdmCouplingExp;
expArray(end+1) = mdmWeightBankExp;
expArray(end+1) = mdmWeightBankSafeExp;
expArray(end+1) = busNtwrkExp;
expArray(end+1) = busNtwrkWDMExp;
expArray(end+1) = busNtwrkInvExp;


% Define how they are put together in the topcell
% Currently, just a name is needed
mirrorNets = ones(1,contCounter);
multiInfo = struct('name','topcell', ...
	'putNextTo',{{'mdmCouplingExp','mdmWeightBankExp'}}, ...
    'rotate',{{'busNtwrkExp'}}, ...
	'L1netMirror',mirrorNets, ...
	'centering',false);

%createMasterHDFfromTypes('mytestfile.h5',expArray);
hdfFilename = [cad.name, '.h5'];
%hdfFilename = [];

% call functions to build all GDS files
delete('./*.gds'); delete('./Cells/*');
buildInfo = BuildCellsFromInfo(expArray,hdfFilename);
BuildMultiCell(buildInfo, multiInfo);
addpath('./Cells');

sprintf('Merging final files. If it crashes, <a href="matlab: opentoline(%s,25)">go here</a>!', which('MergeGDS.m'))
MergeCells;
delete('cadMeta.mat');
\end{lstlisting}
\subsection{Mach-Zehnder Interferometer}
\begin{lstlisting}
%% device_MultiInterferometer
% Author : Ethan
% param format: [coupling length; multi-mode fiber width]
function [topcell, infoOut] = device_MultiInterferometer(topcell, infoOut, phW, len, param)

% Fetch Parameters
cLen    = param(1);
mmWidth = param(2);

% Constants (um)
taperLen    = 50;
arcLen      = 10;
vertSep     = 100;
interLen    = 120;

cplSep      = 0.5;

% Length Check
minLen = 2*taperLen + 2*cLen + 8*arcLen + interLen;
if len < minLen
    [topcell,infoOut] = PlaceRect(topcell,infoOut,len,phW.w,phW.layer,phW.dtype);
    return
end

% Calculate Padding
padLen = (len - minLen) / 2.0;

% Approach
[topcell,infoOut] = PlaceRect(topcell,infoOut,taperLen,phW.w,phW.layer,phW.dtype);
[topcell, infoOut] = PlaceArc(topcell, infoOut, 90, arcLen, phW.w, phW.layer, phW.dtype);
[topcell,infoOut] = PlaceRect(topcell,infoOut,vertSep,phW.w,phW.layer,phW.dtype);
[topcell, infoOut] = PlaceArc(topcell, infoOut, -90, arcLen, phW.w, phW.layer, phW.dtype);

%%% Multi-Mode Fiber %%%
infoOutMM = infoOut;
if infoOutMM.ori == 0
    infoOutMM.pos(2) = infoOut.pos(2) + phW.w / 2.0 + cplSep + mmWidth / 2.0;
    infoOutMM.pos(1) = infoOut.pos(1) - arcLen - taperLen;
end
if infoOutMM.ori == -90
    infoOutMM.pos(2) = infoOut.pos(2) + arcLen + taperLen;
    infoOutMM.pos(1) = infoOut.pos(1) + phW.w / 2.0 + cplSep + mmWidth / 2.0;
end

[topcell,infoOutMM] = PlaceRect(topcell,infoOutMM,taperLen,0.001,phW.layer,phW.dtype, 'endwidth', mmWidth);
[topcell,infoOutMM] = PlaceRect(topcell,infoOutMM,6*arcLen + 2*cLen + 2*padLen + interLen,mmWidth,phW.layer,phW.dtype);
[topcell,~]         = PlaceRect(topcell,infoOutMM,taperLen,mmWidth,phW.layer,phW.dtype, 'endwidth', 0.001);

%%% End Multi-Mode Fiber %%%

% Coupler
[topcell,infoOut] = PlaceRect(topcell,infoOut,cLen,phW.w,phW.layer,phW.dtype);

% Detatch
[topcell, infoOut] = PlaceArc(topcell, infoOut, -90, arcLen, phW.w, phW.layer, phW.dtype);
[topcell,infoOut] = PlaceRect(topcell,infoOut,vertSep,phW.w,phW.layer,phW.dtype);
[topcell, infoOut] = PlaceArc(topcell, infoOut, 90, arcLen, phW.w, phW.layer, phW.dtype);

% Padding and Interleaver
[topcell,infoOut] = PlaceRect(topcell,infoOut,padLen,phW.w,phW.layer,phW.dtype);
for i = 1:3
    [topcell, infoOut] = PlaceArc(topcell, infoOut, 90, arcLen, phW.w, phW.layer, phW.dtype);
    [topcell,infoOut] = PlaceRect(topcell,infoOut,vertSep/2,phW.w,phW.layer,phW.dtype);
    [topcell, infoOut] = PlaceArc(topcell, infoOut, -90, arcLen, phW.w, phW.layer, phW.dtype);
    [topcell, infoOut] = PlaceArc(topcell, infoOut, -90, arcLen, phW.w, phW.layer, phW.dtype);
    [topcell,infoOut] = PlaceRect(topcell,infoOut,vertSep/2,phW.w,phW.layer,phW.dtype);
    [topcell, infoOut] = PlaceArc(topcell, infoOut, 90, arcLen, phW.w, phW.layer, phW.dtype);
end
[topcell,infoOut] = PlaceRect(topcell,infoOut,padLen,phW.w,phW.layer,phW.dtype);

% Second Approach
[topcell, infoOut] = PlaceArc(topcell, infoOut, 90, arcLen, phW.w, phW.layer, phW.dtype);
[topcell,infoOut] = PlaceRect(topcell,infoOut,vertSep,phW.w,phW.layer,phW.dtype);
[topcell, infoOut] = PlaceArc(topcell, infoOut, -90, arcLen, phW.w, phW.layer, phW.dtype);

% Second Coupler
[topcell,infoOut] = PlaceRect(topcell,infoOut,cLen,phW.w,phW.layer,phW.dtype);

% Second Detatch
[topcell, infoOut] = PlaceArc(topcell, infoOut, -90, arcLen, phW.w, phW.layer, phW.dtype);
[topcell,infoOut] = PlaceRect(topcell,infoOut,vertSep,phW.w,phW.layer,phW.dtype);
[topcell, infoOut] = PlaceArc(topcell, infoOut, 90, arcLen, phW.w, phW.layer, phW.dtype);
[topcell,infoOut] = PlaceRect(topcell,infoOut,taperLen,phW.w,phW.layer,phW.dtype);

end
\end{lstlisting}
\subsection{MDM/WDM Weight Bank}
\begin{lstlisting}
%% device_MDMWeightBank
% Author : Ethan
% param format: [coupling length; multi-mode fiber width]
function [topcell, infoOut, infoMetalOut] = device_MDMWeightBank(topcell, infoIn, phW, MeT, param, varargin)
infoMetalOut = {};

% Fetch Parameters
numModes    = param(1);

mdmFile = matfile('../Functions - AlexUtil/MDMparams.mat','Writable',false);
mmWids = mdmFile.wuse;
taperLen = mdmFile.taper;

% Get a filter bank, or make one
if isempty(varargin)
	numWL       = param(2);
	mrr = Microring(.2,phW.layer,phW.dtype,'w',mmWids(1),'radius',phW.r,'straightLength',2);
	fbObj = FilterBank_v2(Waveguide(.5,[1;0],10,10),Wire(4,[11;0],10,20),mrr,'orderN',numWL,'dL',.1,'busHold',8);
else
	fbObj = varargin{1};
end

% Is it hooked on the drop port
if length(varargin) < 2
	ishooked = 2;
else
	ishooked = varargin{2};
end

if length(varargin) < 3
	side = 1;
else
	side = varargin{3};
end


% Draw Approach
[topcell,infoIn] = PlaceRect(topcell, infoIn, phW.r, mmWids(numModes), phW.layer, phW.dtype);

% Draw Filter Bank
n = numModes;
infoThru = infoIn;
infoOut = 0;
mmWids(end+1) = mmWids(end);
%radius = 10;

prevInfoCross = [];

while n > 0
    
    % WDM Filter
    fbObj.busWGwidth = [mmWids(n); mmWids(n)];
    fbObj.meT = MeT;
    fbObj.recalculate();
    [topcell, infoThru, infoCross, infoContra, infoMet] = fbObj.place(topcell,infoThru,'side',side);
    if n == numModes
        infoWrap = infoContra;
    end
    
    % Draw Drop Demux, or if it's the last one draw a terminator
    if n > 1
        demux = ModeMUX(phW,[n,n-1,side]);
        [topcell, infoThru, ~] = demux.place(topcell, infoThru, side);
	else
		% Kill taper
		[topcell, ~] = PlaceRect(topcell, infoCross, taperLen, mmWids(1), phW.layer, phW.dtype,'endwidth',0.01);
    end
    
    % Draw Add Mux
    if n == numModes && numModes > 1
		magic1 = 30;
        infoMuxOut = SplitInfo(infoThru, 1);
        [topcell,infoMuxOut] = PlaceArc(topcell, infoMuxOut, -90*side, phW.r, phW.w, phW.layer, phW.dtype);
        [topcell,infoMuxOut] = PlaceRect(topcell, infoMuxOut, magic1, phW.w, phW.layer, phW.dtype);
        [topcell,infoOut] = PlaceArc(topcell, infoMuxOut, 90*side, phW.r, phW.w, phW.layer, phW.dtype);
	elseif numModes > 1
        if n > 1
            infoMuxOut = SplitInfo(infoThru, 1);
        else
            infoMuxOut = infoThru;
        end
        [topcell,infoOut] = PlaceRect(topcell, infoOut, abs(infoOut.pos(2) - infoMuxOut.pos(2)), mmWids(numModes-n), phW.layer, phW.dtype);
        mux = ModeMUX(phW, [numModes - n, numModes - n + 1, -side]);
        combInfo = MergeInfo(infoOut, infoMuxOut);
		if side == -1, combInfo.pos = flipud(combInfo.pos); end
        [topcell, infoOut] = mux.place(topcell, combInfo);
	else
		infoMuxOut = infoThru;
		[topcell,infoOut] = PlaceRect(topcell, infoMuxOut, taperLen + phW.sp, mmWids(1), phW.layer, phW.dtype);
    end
    
    if n < numModes
       % Connect infoContra and PrevInfoCross
       [topcell, ~] = PlaceSBend(topcell, prevInfoCross, abs(prevInfoCross.pos(2) - infoContra.pos(2)), infoContra.pos(1) - prevInfoCross.pos(1), phW.r, mmWids(n), phW.layer, phW.dtype);
    end
    
    % Draw Taper, new previnfocross
    if n > 1
        infoThru   = SplitInfo(infoThru, 2);
        [topcell, prevInfoCross] = PlaceRect(topcell, infoCross, taperLen, mmWids(n), phW.layer, phW.dtype,'endwidth',mmWids(n-1));
	end
	
	% Accumulate metals
	if isempty(infoMetalOut)
		infoMetalOut = infoMet;
	elseif ~isempty(infoMetalOut{1}) && ~isempty(infoMet{1})
		infoMetalOut = {MergeInfo(infoMetalOut{1},infoMet{1}), MergeInfo(infoMetalOut{2},infoMet{2})};
	end
	
    n = n - 1;
end

% Draw Drop Wrap-Around
if ishooked > 0
	[topcell,infoWrap] = PlaceArc(topcell, infoWrap, -180*side, phW.r, mmWids(numModes), phW.layer, phW.dtype);
end
if ishooked > 1
	[topcell,infoWrap] = PlaceRect(topcell, infoWrap, abs(infoWrap.pos(2) - infoOut.pos(2)), mmWids(numModes), phW.layer, phW.dtype);
end

infoOut = MergeInfo(infoOut, infoWrap);


end
\end{lstlisting}
\subsection{Folded Bus Network}
\begin{lstlisting}
%% device_NtwrkArr
% Author : Ethan

% Dummy device for modular type experiments (which require a certain len)
% See device_blank for arrayed type experiments
function [topcell, infoAll, infoMetalOut, refs] = device_NtwrkArr(topcell, infoAll, phW, MeT, params, varargin)
infoMetalOut = [];
refs = [];
deviceModel = varargin{1};
dParams     = varargin{2};
numModes    = dParams(1);
ioNum = size(infoAll.pos,1);

% Import Params
mdmFile = matfile('../Functions - AlexUtil/MDMparams.mat','Writable',false);
mmWids = mdmFile.wuse;
taperLen = mdmFile.taper;

% Get a microring, or make one
if isempty(varargin)
	%numWL       = param(2);
	mrrObj = Microring(.2,phW.layer,phW.dtype,'w',mmWids(1),'radius',phW.r,'straightLength',2);
else
	mrrObj = varargin{1};
end

% Construct Demux
demux  = ModeMUX(phW,[numModes,1,1]);

% Draw Modulator
mrr = Microring(.2,1,0,'w',mmWids(1),'radius',10,'straightLength',2);
fb = FilterBank_v2(Waveguide(.5,[1;0],10,10),Wire(4,[11;0],10,20),mrr,'orderN',1,'dL',.1,'busHold',8);
fb.busWGwidth = [mmWids(1); mmWids(1)];
fb.meT = MeT;
fb.recalculate();

n = numModes;
prevInfoCross = [];
while n > 0
    fbObj = fb.copy();
    
    infoPwr = SplitInfo(infoAll, ioNum-numModes+n);

    % Approach
    [topcell,infoPwr] = PlaceRect(topcell, infoPwr, taperLen, mmWids(1), phW.layer, phW.dtype);
    [topcell,infoPwr] = PlaceArc(topcell, infoPwr, -90, phW.r, mmWids(1), phW.layer, phW.dtype);
    
    % Modulator
    fbObj.busWGwidth = [mmWids(1); mmWids(n)];
    fbObj.recalculate();
    
    [topcell, infoThru, infoCross, infoContra, ~] = fbObj.place(topcell,infoPwr,'side',1);
    
    % Kill Taper
    [topcell, ~] = PlaceRect(topcell, infoThru, taperLen, mmWids(1), phW.layer, phW.dtype,'endwidth',0.01);
    
    if n > 1
        % Adiabatic Taper
        [topcell, prevInfoCross] = PlaceRect(topcell, infoCross, taperLen, mmWids(n), phW.layer, phW.dtype,'endwidth',mmWids(n-1));
    end
    
    if n < numModes
        % S bend to meet taper
        [topcell, ~] = PlaceSBend(topcell, infoContra, abs(prevInfoCross.pos(2) - infoContra.pos(2)), infoContra.pos(1) - prevInfoCross.pos(1), phW.r, mmWids(n), phW.layer, phW.dtype);
    else
        % Contra is Output!
        infoOut = infoContra;
    end
    
    n = n - 1;
end

% Draw Hairpin to Dendrites
[topcell, infoOut] = PlaceArc(topcell, infoOut, -90, phW.r, mmWids(numModes), phW.layer, phW.dtype);
[topcell, infoOut] = PlaceRect(topcell, infoOut, taperLen * 5, mmWids(numModes), phW.layer, phW.dtype);
[topcell, infoOut] = PlaceArc(topcell, infoOut, -90, phW.r, mmWids(numModes), phW.layer, phW.dtype);

% Draw Drop / Continue
[topcell, infoOut] = deviceModel(topcell, infoOut, phW, MeT, dParams);

infoWB = SplitInfo(infoOut, 1);

infoCont = SplitInfo(infoOut, 2);

% Draw Weight Bank 1
[topcell, infoOut] = deviceModel(topcell, infoWB, phW, MeT, dParams);
infoMid = SplitInfo(infoOut, 1);
infoBot = SplitInfo(infoOut, 2);

% Place Unapproach
[topcell, infoMid] = PlaceArc(topcell, infoMid, -90, phW.r, mmWids(numModes), phW.layer, phW.dtype);

[topcell, infoBot] = PlaceRect(topcell, infoBot, 50, mmWids(numModes), phW.layer, phW.dtype);
[topcell, infoBot] = PlaceArc(topcell, infoBot, -90, phW.r, mmWids(numModes), phW.layer, phW.dtype);
[topcell, infoBot] = PlaceRect(topcell, infoBot, infoBot.pos(1) - infoMid.pos(1), mmWids(numModes), phW.layer, phW.dtype);


% Place Demux
[topcell, infoMid] = demux.place(topcell, infoMid);
[topcell, infoBot] = demux.place(topcell, infoBot);

infoComb = MergeInfo(infoMid, infoBot);

% Align Outputs

infoNewComb = 0;
for n = 1:2*numModes
    infoTest = SplitInfo(infoComb, n);
    [topcell, infoTest] = PlaceRect(topcell, infoTest, infoTest.pos(1) - infoAll.pos(1, 1), phW.w, phW.layer, phW.dtype);
    if isstruct(infoNewComb)
        infoNewComb = MergeInfo(infoNewComb, infoTest);
    else
        infoNewComb = infoTest;
    end
end

% Draw Weight Bank 2
[topcell, infoCont] = PlaceRect(topcell, infoCont, 100, mmWids(numModes), phW.layer, phW.dtype);
[topcell, infoCont] = PlaceArc(topcell, infoCont, 90, phW.r, mmWids(numModes), phW.layer, phW.dtype);
[topcell, infoCont] = PlaceArc(topcell, infoCont, -90, phW.r, mmWids(numModes), phW.layer, phW.dtype);
[topcell, infoOut] = deviceModel(topcell, infoCont, phW, MeT, dParams);
infoMid = SplitInfo(infoOut, 1);
infoBot = SplitInfo(infoOut, 2);

% Place Unapproach
[topcell, infoMid] = PlaceArc(topcell, infoMid, -90, phW.r, mmWids(numModes), phW.layer, phW.dtype);

[topcell, infoBot] = PlaceRect(topcell, infoBot, 50, mmWids(numModes), phW.layer, phW.dtype);
[topcell, infoBot] = PlaceArc(topcell, infoBot, -90, phW.r, mmWids(numModes), phW.layer, phW.dtype);
[topcell, infoBot] = PlaceRect(topcell, infoBot, infoBot.pos(1) - infoMid.pos(1), mmWids(numModes), phW.layer, phW.dtype);


% Place Demux
[topcell, infoMid] = demux.place(topcell, infoMid);
[topcell, infoBot] = demux.place(topcell, infoBot);

infoComb = MergeInfo(infoMid, infoBot);

% Align Outputs

infoNewComb2 = 0;
for n = 1:2*numModes
    infoTest = SplitInfo(infoComb, n);
    [topcell, infoTest] = PlaceRect(topcell, infoTest, infoTest.pos(1) - infoAll.pos(1, 1), phW.w, phW.layer, phW.dtype);
    if isstruct(infoNewComb2)
        infoNewComb2 = MergeInfo(infoNewComb2, infoTest);
    else
        infoNewComb2 = infoTest;
    end
end

% Finalize
infoNewComb2 = MergeInfo(infoNewComb, infoNewComb2);
infoTop  = SplitInfo(infoAll, (ioNum - numModes + 1):ioNum );
infoAll = MergeInfo(infoTop, infoNewComb2);
end
\end{lstlisting}

\end{document}